\hsize=155mm  \vsize=215mm
\hoffset=0mm
\baselineskip 15 pt
\overfullrule 0pt
\magnification=\magstep1
\font\mi=cmbx10 scaled\magstep1
\overfullrule 0pt              
\def \zer {^  {(0)}    }
 \def \noi {\noindent}
 \def \>  { \rightarrow }
 \def \alp {\alpha } 
  \def \lie  { {\cal L} }
\def \<  { \leftarrow }
 \def  \pat {\widehat p}
\def \qat  {\widehat q}
 \def  \lich {[1]}    \def  \hak {[2]}      \def  \isrkan {[3]}   
     \def  \fw {[4]}      \def  \vdw {[5]}        \def  \aadis {[6]}          
   \def  \dvlxx {[7]}
 \def  \dvlxxv {[8]}       \def  \hakII {[9]}     \def  \belpredic {[10]}
 \def  \lapsan {[11]}      \def  \horpir {[12]}      \def  \horstat {[13]}  
\def  \berg {[14]}    \def  \abr {[15]}      
 \def  \aren {[16]}             \def  \dvlxxvi {[17]} 
 \def  \dvlxxvii {[18]}       \def  \lus {[19]}       \def  \constr {[20]}      
 \def  \byac {[21]}       \def  \hunz {[22]}       \def  \dvscat {[23]}      
 \def  \horrohr  {[24]}      
   \def  \cur {[25]}   
   \def  \dvqu {[26]}   \def  \belfus {[27]}   
 \def  \hj {[28]}        \def  \lapbar {[29]}

 \centerline {\mi  Direct Interactions}
\centerline {\mi in Relativistic Statistical Mechanics}

\bigskip
\centerline {\bf Philippe Droz-Vincent }
\bigskip
\centerline {Gravitation et Cosmologie Relativistes }
\centerline   {C.N.R.S. URA 769, Universit\'e Pierre et Marie Curie}
\centerline  {Tour 22-12, boite courrier 142}
\centerline  {4 place Jussieu 75252 Paris Cedex 05, France}
\bigskip
\vfill        
\bigskip
\noi   {\it  Directly interacting particles are considered in the multitime  
formalism of predictive relativistic mechanics.
When the equations of motion leave a phase-space volume invariant, it turns 
out that the phase average of any first integral, covariantly defined as a flux 
across a $7n$-dimensional surface, is conserved.
The Hamiltonian case is discussed, a class of simple  models is exhibited, 
and a tentative definition of equilibrium is proposed}.


\medskip\vfill\eject
\noi   {\bf 1. Introduction}

\noi   It is a pleasure to contribute this paper in the honor of 
prof. Larry P. Horwitz who has done (and keeps doing !) so much in 
practically all the domains of relativistic dynamics.  

  Relativistic statistical mechanics has a long story, 
but we may notice that, whereas the theory of relativistic {\it ideal gases\/}
 has received  deep and   detailed developments,      little has been 
achieved in order to  account for   {\it mutual  interactions\/}  beetwen 
particles.
 Most work available in the litterature concerns kinetic theory {\lich},
  and it 
is only through wild simplifications  that collision terms can be somehow
 ``derived" from the underlying  $n$-body dynamics.
For instance  various  results are obtained  with the help of a
 straight-line   approximation 
 which amounts to calculate the  force applied to each particle  as if all 
other  particles were undergoing free motion. 
This treatment retains  very little information (if any) about  the underlying
 dynamics.

\noi
Another approach consists in giving up manifest covariance in 
order to deal with a more familiar framework, where the concepts and methods
 of newtonian statistical mechanics are intensively  employed.
This approach is plagued with unpleasant complications (and sometimes serious  
difficulties)  concerning relativistic invariance.

\noi
Looking for a more satisfactory   theory, where interactions between 
particles would be described in  a fully covariant fashion,
 several authors  {\hak}{\isrkan}             have hesitated 
between two  points  of view: field theory  and   action at a distance.

\noi
In fact, if one  aims at rigorous results, it is hardly possible 
 to deal with the tremendous infinity of the field degrees of freedom.
It seems more convenient to eliminate the field variables, and to formulate 
 equations of motion for the  {\it particle degrees of freedom\/} only.

\noi
This is the case for  Feynman-Wheeler electrodynamics
 {\fw}, or more generally 
for any interaction {\it a la\/} Van Dam and Wigner {\vdw}.
But these  versions  of relativistic dynamics  are not yet  tractable;
in the best cases, they lead to difference-differential equations (they involve 
retarded  arguments) and seem to require an infinite number of 
initial data for determining the trajectories.

\noi
In contradistinction, the modern theories of   {\it  directly   interacting 
particles\/} \aadis are tailored for dealing with
 {\it only a finite number of degrees of freedom\/}.
They naturally suppose that the field  degrees of freedom have been 
eliminated,
but in addition they assume  from the start that a finite number of
 initial data is sufficient in    order to determine the evolution of the 
system.
  In this sense they are a simplification of the approach initiated by 
 the authors  of refs {\fw}{\vdw}. 

\noi
In this spirit,   we  have proposed many years ago that an
 $n$-body relativistic system be described by truly differential equations 
of motion, involving $n$ independent evolution parameters {\dvlxx}{\dvlxxv}.
Our  departure from the Feynman-Wheeler scheme is the requirement that the 
right-hand sides of the equations depend only on the   positions 
(in space-time) and their  first derivatives {\hakII}.
Following the specialized litterature, let us refer to this approach as 
{\sl predictive mechanics}; clearly this terminology accounts for the
  possibility to  determine, in principle, world lines from the knowlege of
 (a finite number of)   initial data {\belpredic}.
 
\noi
Lapiedra et al. \lapsan 
have undertaken  an  application of predictive methods to 
relativistic statistical mechanics.
However  their definitions for  averages, normalization and 
equilibrium are not manifestly covariant.

\noi
A quite different view of direct relativistic interactions has been proposed 
by   Horwitz and  Piron {\horpir};
  in their approach, a unique "historical time" is supposed to  
describe the motion  of any number of particles; in other words a 
single-parameter evolution group acts in a finite-dimensional covariant 
phase space.
Along this line, Horwitz {\it et al.\/} \horstat have extensively developed a 
"statistics of events" considered off the mass shell.

\noi
In contradistinction we  intend to build a 
{\it  statistics of orbits\/}, in agreement with Hakim's philosophy (which is 
after all very close to the principles of predictive mechanics).

\bigskip
\noi
In a covariant framework like ours, 
 the relativistic analog of a  function
 integrated over phase space usually  appears  as the flux of
 a generalized current   accross some manifold {\berg} {\hak}.
 For instance the $n$-body distribution function $N$ is normalized through
 the formula  
 $ \int N \alp  =1 $, where $\alp $ is, in our notation,
 a $7n$-differential form introduced by 
 Hakim ($N \alp $ is interpreted as the {\sl numerical flux of particles}).
 Independence of the result with respect to the manifold where integration is 
 performed crucially requires that the integrand be a closed form in the 
 sense of differential geometry.

\noi
 In fact the importance of having $ d(N \alp) = 0 $ was clearly emphasized in 
 ref. {\hak}.
 But in that work  it was assumed from the outset that 
 $ \displaystyle \int _ \Delta  N \alp  $  represents the probability that the 
 system be in a state characterized by a point of phase space belonging to a 
 domain  $ \Delta   $ in  a  $7n $-manifold   $S^ {7n}$ which extends 
to infinity.
Then, invoking that the number of particles is conserved, it was 
{\it concluded\/} that $ d(N \alp) $ is necessarily zero.

\noi
We shall look at these matters in a  different way.
Let us consider  a fixed number $n$ of particles, and suppose that their
 dynamics is given; for instance we write some equations of motion.
Now  we cannot just {\it assume\/} that 
  $ \int _ \Delta N \alp $  is the above  mentioned probability,
 without first {\it proving\/} that
its value does not depend on the cutting surface               $S^ {7n}$
if the same orbits are considered.
In other words the definition of "numerical flux of particles" 
should be consistent with the dynamics: 
  {\it its conservation should stem from the equations of motion\/}
(but we cannot  hope to satisfy this requirement without specifying 
 some technical information  about the nature of these equations).

\noi
The main goal of this article consists in showing that direct interactions, 
especially when formulated as in predictive mechanics, 
 exactly satisfy this requirement.

\bigskip
\noi
Owing to the advantage of dealing with a finite number of degrees of freedom,
a rather rigorous setting can be elaborated. 
We first recollect in Section 2 some useful results of differential geometry 
concerning   measure-preserving vector fields which have vanishing 
 Lie brackets  among themselves
(in these matters we try to speak of incompressible vector fields only,
 postponing  as far as possible the use of Hamiltonian vector fields).

\noi
Section 3 is devoted to the treatment of direct interactions in the 
predictive formalism.
The basic equations of statistical mechanics are introduced following Hakim's 
early work {\hak}.
Then conservation of phase averages easily comes out by straightforward
 application  of the geometrical results gathered in Section 2.
Not only the flux of particles but also the entropy flux is a closed 
differential form.

\noi
The Hamiltonian formalism is considered  in Section 4; a possible ambiguity 
related with the noncanonical nature of physical positions is discussed and 
removed with the help of a certain idea of the underlying quantum mechanics.
Energy,  linear and angular momenta  are introduced  therein.

\noi
A possible definition of  equilibrium ensembles is tentatively 
proposed in Section 5, and the last Section  contains a few concluding remarks.

\bigskip
\noi
 {\bf 2. Geometrical considerations.}

 \noi
In this section we consider an $m$-dimensional orientable manifold  $V^ m$.
Capital indices  $A,B  = 1, ....m $.

\noi
According to the standard notations of differential geometry,  $i_X$ denotes 
the {\sl inner product} (of a differential form) by the vector-field $X$, and 
$\lie _X $ is the Lie derivative.

\noi

 \noi
 In {\it arbitrary\/} coordinates $y^ 1, ...y^ {m} $, a  {\sl volume element}
 is a scalar 
$m$-form  $\eta =  \gamma (y) \   d^ {m} y $, where $\gamma $ is strictly 
positive and transforms as   a scalar density.
If $ \displaystyle\varepsilon _{A_1 , ...A_{m}} $ is the Levi-Civita object, 
                             we define the completely skew-symmetric tensor
$$ \eta _{A_1....A_{m}} = 
                 \gamma   \      \varepsilon _{A_1 .....A_{m} }    $$
Naturally we have set    
$$   d ^ {m} y   = 
{ 1  \over  (m ) ! }\   \varepsilon  _ {A_1  ....A_ {m} }
                             \   dy^ {A_1} \wedge .... dy^ {A_{m}} =
dy^ 1 \wedge  .....  dy^ m   $$
We say that $X$ preserves $\eta$ or equivalently 
that $X$ is {\sl incompressible} when $\lie _X \eta =0$.

\noi
For   $n$ such that $1 \leq n  <m $ 
consider independent vector fields 
$ X_1, ...X_n $. We associate to them   the $(m-n)$-form 
 $$  \alpha _ {(X_a, \eta)} =
 X_1 ^ {A_1} ...X_n ^ {A_n} \     d \Sigma _{A_1....A_n}   \eqno (2.1)    $$
where  we define   
 $$         d \Sigma _{A_1....A_n} =  {1 \over (m-n)!}
\    \eta _{A_1,...A_n A_{n+1}...A_m }  \   
dy^ {A_{n+1}}  \wedge ...\wedge dy ^ {A_m}      \eqno (2.2)  $$
Actually $d \Sigma_ {A_1....A_n} $
is the element of $(m-n)$-surface (imbedded in $V^{8n}$) induced by $\eta$.
 
\noi
Indices $a,b = 1, ...n$. 
Our notation for $\alp$ means to recall that $\alp$ depends not only on 
the vector fields   $X_a$, but also on the volume form.

\medskip
\noi
{\bf Remark}: If all the vector fields $X_a$ are tangent to a $(m-n)$-manifold
$  S^ {m-n} $, then $\alp$ vanishes on this manifold.
 
\medskip
\noi   For typographical convenience we define
$ \displaystyle   i_a =  i_{X_a},   \quad   \lie _a = \lie _{X_a} $.

\noi  In this notation   $ \alp $ is obviously proportional to 
 $ i_ 1 i_ 2 ....i_ n   \  \eta     $, which motivates our attention 
to the statements presented below.

\noi 
 For applications to statistical mechanics it is of interest that $\alp$ be a 
closed differential form (that is $d \alp =0$).  Actually this property 
is ensured by a couple of results presented below.

   \bigskip
\noi
{\bf Proposition I}

\noi
{\sl For  $ n<m$ ,  consider on $V^ m$ the  vector fields
$ X _1, X_2  ...X_n $ ,   preserving the volume element $\eta $.
If  $[   X_a , X_b   ]  =0 $ then  for all   $ r \leq n $   and for all 
$b$
we have }
 $$  \lie _b  (i_1 ....i_r  \  \eta )  =0         \eqno (2.3)  $$

Proof

\noi
The vanishing of Lie brackets implies \abr that $\lie_a$ commutes with $i_b$. 
Hence 
$ \lie _a    i_b \eta =  i_b  \lie _a   \eta  = 0  $.
 Thus $\lie_b  $ commutes with all the contractions $ i_1 ....i_r $.
We end up with   $ i_1 .....i_r \  \lie _b \eta $, obviously zero.

\medskip
\noi
We can also  assert the following  statement
 
\noi
{\bf Proposition II}

\noi
{\sl  Under the same assumptions as above, for all $r \leq n,  \quad
 i_1 .....i_r  \  \eta   \   $ is a closed form}.

\noi
The proof  is by induction.   $ d i_a \eta $ is zero because it coincides with 
$ \lie _a   \eta $. 
If the statement is true for $r-1$, write  the identity 
$$ d \   i_1 ....i_r  \   \eta  =
  \lie _1 (i_2 ....i_r  \eta )  -     i_1 \ d (i_2 ....i_r \  \eta)         $$
In   the second term of   the r.h.s. arises   the differential of 
 $  i_2 ...i_r \eta $, which involves  $r-1 $ inner products and is 
   a closed form because the statement applies to $r-1$.
 Only the first term of the r.h.s. might be nonzero.
One is left with 
$$ d (i_1.....i_r \eta) =  \lie _1  (i_2 .....i_r \eta) $$
but this term also is zero  according to Proposition I.

\medskip
\noi
Since  $\alp $ is proportional to $i_1 ....i_n \eta $, Proposition II 
entails that   $d \alp $ vanishes as announced.

\medskip
\noi
{\bf Remark} 
 The reader who is not familiar with exterior differential calculus can get 
convinced that  $ d \alp $ vanishes with the help of {\it local\/}
 coordinates such   that      $ X_a = \partial  / \partial  y ^ a $  
(notice that in general $ \gamma   \not= 1 $ in these coordinates).
Obviously   $ X^ A _a =  \delta ^ A _a  $, thus 
    $ X^ {A_1}  _a =  \delta ^ {A_1}  _a  $, etc.
Inserting   these values into $ \alpha $ one gets 
$$    \alpha =  {1\over  (m-n)!} \   
 \eta _{1 2....n A_{n+1} ....A_m }  \   
 dy ^ {A_{n+1}}  \wedge .... \wedge  \  dy ^ {A_m}  \eqno  (2.4) $$
valid  only in these particular coordinates.

\bigskip
\noi
With the purpose of applications to averages of various constants of motion,
we make this statement 

\noi
{\bf Proposition III}

\noi
{\sl Under the same assumptions, consider a function $f$ invariant under all 
the vector fields $X$,  that is $ \lie _a   f =0 \  \forall a $.  
We claim that  $ d (f \alp ) =0 $}.
    
\noi
In local coordinates where $ X _a =  \partial / \partial y ^ a $ it is easy to 
check that $ df \wedge \alp = 0$.  But a global statement requires the 
following calculations.

\noi
Owing to the well-known  identity 
$ i_X (A \wedge B) \equiv  i_X A \wedge B +  (-)^ k A \wedge i_X B $
 we notice this 

\noi {\sl Lemma

\noi
If $i_a B $ vanishes for all $a = 1,2 ....n$, then $\forall   r \leq n$},
$$ (i_1 .... i_r  A) \wedge  B  = i_1 ....i_r (A \wedge B)    \eqno (2.5) $$
easily proved by induction.

\noi
Apply it to  $A = \eta,   \quad  B=df $ where  $ \lie _a f =0 $.
Since  degree of  $\eta $ is maximal, $\eta \wedge df =0$, which proves that 
$df  \wedge \alp =0$, hence $ d (f \alpha) = 0 $.

\bigskip
\noi

\medskip
\noi
This result will receive several applications in the context of  relativistic 
statistical mechanics:
normalization of the  distribution function, intrinsic definition of entropy 
flux, of average (linear or angular) momentum.

\bigskip
\noi
{\bf 2.1.  Rescaling}

\noi  
In connection with the problem of reparametrization of world lines, another 
property of  differential geometry is worth noticing.

\noi
{\bf  Definition}.

\noi
{\sl Let $X_1, ....X_n $ have mutually vanishing Lie brackets.
An abelian   rescaling  is  the replacement of these vector fields by 
$$ Y_1 = \phi _1 X_1, \quad .....Y_n = \phi _n   X_n   $$
where  $\phi_1,...\phi_n$ are strictly positive scalars satisfying the 
condition
$ \lie _a  \phi _b = 0 $ for $a \not= b $}.

\noi
We can obviously check that {\sl the new vector fields $Y_a$ have mutually
 vanishing Lie brackets among themselves}, which justifies our terminology.
 More general rescalings of the $X_a$'s  would spoil the important Lie 
bracket condition and will not be considered in this paper.

\medskip
\noi
{\bf Proposition IV} 

\noi
{\sl Let  $\eta $ be  a volume invariant by the vector fields 
$X_1, ...X_n$ with mutually vanishing Lie brackets.
Then  $ (\phi _1 ....\phi _n )^ {-1}  \eta $
 is a volume invariant by the rescaled   vector fields $Y_1, ...Y_n$}.

\noi
The proof is straightforward:
 Applied to any form of maximal degree, like the volume element,
  $\lie _a $ is given by $ di_a $.        Notice that
$$ i_{\phi_1 X_1}...i_{\phi_n X_n} 
(\phi_1 ^ {-1} ...\phi_n ^ {-1})  \eta = i _{X_1} ....i_{X_n} \eta $$
and differentiate.

\noi
Now we immediately get 

\noi {\bf Proposition V}

\noi
{\sl Under the assumptions of Proposition IV, the $(m-n)$-form $\alp $ 
associated with the vector fields $X_a$ and  
the volume element $\eta $ is invariant under the replacement 
$ X_a  \>  \phi_a  X_a , \qquad  
 \eta \> (\phi _1 ....\phi _n)^ {-1} \  \eta $}.


\bigskip
\noi
  {\bf 3. Statistical mechanics of predictive $n$-body systems}.

\noi
A predictive system of $n$ particles, labelled by subscripts $a,b,...$ 
running from $1$ to $n$,  
can  be defined by the second order {\it differential\/} equations  \dvlxx
$$  {d^ 2 x_a ^ \alp  \over  d \tau_a ^ 2}= \xi_a  ^  \alp (x,v) \eqno (3.1) $$
This picture exhibits some analogy with the Feynman-Wheeler equations of 
motion; in particular each world-line is parametrized by its own parameter.
But in contradistinction, the above equations are simply {\it differential\/}
 equations (they {\it do not\/} involve  retarded arguments). 
 In contrast to Galilean mechanics, the r.h.s. of the equations of motion 
cannot be choosen arbitrarily. This is the price paid for a manifestly 
covariant formulation.

\noi
The above system is integrable  and provides world-lines 
$ x_a = x_a (\tau _a)$ provided the right-hand sides satisfy the
 {\sl predictivity conditions}.
This scheme amounts to consider the tangent bundle $(T(M^4))^n$ as phase space.
   Setting  
 $ \displaystyle v_a =  {dx_a  \over d \tau _a}$ we have in the bundle 
"natural coordinates" $x,v$.    
The generators of the evolution group are the vector fields
$$ X_a =  v_a \cdot  {\partial \over \partial x _a } 
+          \xi _a \cdot    {\partial \over \partial v _a }  \eqno (3.2) $$
They can be viewed as  linear homogeneous differential operators 
acting on phase functions;  in this sense they are  Liouville operators.  
(Accordingly we shall often denote 
 $Xf =  {\lie _X }f $ 
the Lie derivative of a function $f$ by  a vector field $X$).     
The lift of world-lines in phase space is formed by the integral curves of 
$ X_1, X_2, ... X_n $.
The predictivity condition can be written in terms of  Lie brackets 
{\dvlxx}{\dvlxxv}
$$  [   X_a , X_b    ] =0    \eqno (3.3) $$
Notice that these conditions are nonlinear in the "accelerations" $\xi_a$.

\noi
The  evolution group  has $n$ parameters and is {\it abelian\/}; its orbits are
 the integral curves of  $X_1, ...X_n$, 
and their projection over configuration space $M^ {4n}$ gives 
the world lines.

\noi
A {\sl first integral}
 is of course a quantity $f(x,v) $ satisfying the relations
$X_a f = 0 ,  \forall a $.
A {\sl partial integral} relative to  $X_a$ is defined by   
$ X_b f = 0,  \forall b \not= a $.

\bigskip
\noi
When the evolution parameters $\tau _a $ are affine parameters 
(the  proper-times respectively divided by the masses),
 the r.h.s. of (3.1)  satisfy the additionnal condition 
   $  v_a  \cdot \xi _a  = 0 $.
In this case we say that our formulation is {\sl autochronous}. When the 
above condition fails to be satisfied, we say that the formulation is 
{\sl heterochronous}. In this case $v_a ^ 2  $ are not anymore constants of 
the motion, but they are still required to remain strictly positive.

\noi
Some care is needed however, not only in view of possible ambiguities,
discussed later on, but also in order to maintain timelikeness of world lines.
As soon as condition $v \cdot \xi $ is relaxed, it becomes necessary to make 
sure that $n$ first integrals, say  $ K_1, ....K_n $ can be identified as 
 half-squared masses.  
    Assuming  that this is actually possible,  the physically relevant part of
  phase space is that piece of $ (T(M^4) ^ n) $ 
which corresponds to positive values of $K_a$.

\noi
Then we can always rescale the Liouville operators to an  autochronous 
formulation, as follows.
Take  $ \phi _a = \sqrt {2K_a  / v_a ^2} $. Easy check that $\lie_a \phi _b $
vanishes for $a \not=  b$, which makes $\phi_a $ admissible for a rescaling 
according to section 2.
If $Y_a  = \phi_a  X _a $,  and  $w_a ^ \alp = \phi _a v_a $, it turns out
that  $ w_a \cdot  w_a =  2K $ hence the square of all $w_a$ is constant. In 
other words, {\it the rescaled formulation is autochronous\/} and we end up with 
$n$ mass-shell constraints  $w_a ^ 2 = m_a ^ 2 $.

\noi
{\sl Remark}

\noi
Since  the Poincar\'e group is implemented in phase space in terms of not 
only positions but also velocities, it is relevant to realize that the 
definition of this group is invariant by abelian rescaling, 
provided that the scaling factors $\phi _a $ are themselves
 {\sl scalar Poincar\'e invariants}.

\noi
Recall this definition : 
Scalar invariants of the Poincar\'e group {\aren}   are arbitrary 
combinations of  all the scalar products made of "vector invariants"  like:
$$ x_a - x_b,    \qquad  v_c,     
    \qquad   (x_a -x_b) \wedge v_c  \wedge v_d       $$

\medskip
\noi
Why did we bother with consideration of  nonaffine parameters?  
 \noi
At first sight it seems natural to impose  affine 
parameters from the outset,
 and our first version of predictive systems \dvlxx was restricted to 
this case. 
But soon it turned out that allowing for arbitrary parametrizations 
 has several advantages {\dvlxxvi}{\dvlxxvii}.     
Indeed  
the use of nonaffine parameters      facilitates  the construction of 
hamiltonian predictive systems in closed form, along the line of the {\it a 
priori hamiltonian approach\/} {\dvlxxvii}.
This procedure was proved  \lus to be equivalent with the constraint approach
 {\constr}; let us stress that  the key for a contact  of  predictive 
mechanics with this   alternative formulation (based upon 
Dirac's constraints theory)  is  an equal-time condition that is in general
 {\it not compatible\/} with the affine parametrization of world lines.
In particular  the most tractable toy model of relativistic 
interaction is a two-body harmonic oscillator formulated in terms of 
parameters which are not affine {\dvlxxv}{\dvlxxvii}.

\bigskip
\noi
The structure determined by the vector fields $X_1,.....X_n$
over phase space allows to  {\it formally\/}  define  the evolution group
through  $ U _{\tau _1,  ....\tau _n} = 
 \exp {(-\tau _1 X_1  .....- \tau _ n X _n)}    $. 

\bigskip
\noi
There is {\it a priori\/} no Liouville theorem at our disposal, but we can still 
look for a conserved volume element, that is an everywhere  positive 
 $8n$-form  invariant by action of all the vector fields $X_a$.
 If such a volume preserved by the motion, say  $\eta $, 
can be  found (that is  $ \lie _a \eta =0$)
   it is generally not unique. 
For a given system of vector fields $X_a$, the invariant volume 
 (if it globally exists) can be determined only   {\it up to a factor\/} 
$ \Lambda $, which is necessarily a positive first integral.
Let us write  
$ \eta = \gamma \  d^{8n}   y $, where $\gamma $ is 
some scalar density in arbitrary coordinates.
Now it is possible to interprete   $U$ as a unitary operator because we have a 
scalar product for phase space functions, 
say  $ <f,g> = \int f^*  g  \    \eta  $
and this product is invariant under 
evolution.

\medskip
\noi    It is not yet necessary to assume a Hamiltonian formalism.
 In fact the contents of  Section 2 imply that several  general results can 
be derived under the simple assumption of a conserved volume element;
  these results seem to open a  way to abstract developments in the spirit
 of mathematical ergodic theory. 
We shall see later  what are the limitations of this point of view.

\bigskip
\noi
 {\bf Asymptotic considerations}

\noi  The interest of referring a system of interacting particles to its
 interaction-free limit has been 
recently emphasized {\byac}. As temporal conditions have been invoked in 
order to determine this procedure, let us sketch how we can figure 
the asymptotic time behaviour of these particles.
It is convenient  to remember that a {\it classical scattering\/}
 theory parallel to the  quantum one is possible, and has been actually 
developed in the newtonian context {\hunz}.
A classical relativistic scattering theory could be considered as well, in 
close analogy to the many-time formulation of quantum relativistic scattering 
theory available in the litterature {\dvscat}{\horrohr}.
Exponentiating the  Liouville-operators we get 
$ U =   \exp {(-\tau _1  X_1   ...... -\tau_n  X_n )}    $.
In a similar way the  system of  $n$ free particles is characterized by the 
Liouville-operators  $X_1 \zer , .....X_n \zer $  where 
$  X_a \zer  =  v_a  \cdot  \partial _a   $.  They correspond to the evolution 
operator   $U \zer =   
     \exp {(-\tau _1  X_1  \zer   ......-\tau _n  X_n \zer )} $.
Then introduce classical relativistic Moeller operators as the limits
$$ \Omega ^ \pm =  \lim  U^ {-1}  U \zer      $$
for  $ \tau _1, ... \tau _n $  alltogether $ \>  \infty $.
{\it If the limits  actually exist\/}  in the strong sense,
 which requires that they are independent of the  order
(as happens in the quantum theory as proved by Horwitz and Rohrlich 
{\horrohr}),   then it is reasonable to expect the intertwining relations
$$     U \Omega ^\pm =  \Omega ^\pm   U \zer ,
\qquad  \quad      X_a  \Omega ^\pm =  \Omega ^\pm   X_a \zer   \eqno (3.4) $$
This gives  the principle of a map between the orbits of the interacting system 
and the solutions of the free-particle motion, invertible map insofar as no 
bound  state is present.
This seems to suport the claim by Ben Ya'acov  that the evolution of the 
distribution function can be fully described in terms of the interaction-free 
limit of  the system {\byac}.
However one must be cautioned that no rigorous result about $\Omega ^ \pm $ 
is by now available, 
and if anything is going to be proved, it will probably concern short-range 
interactions rather than electromagnetism.

\noi
The Liouville " theorem" is essential 
  \footnote *{Quotation marks refer to the fact that, for the 
moment, volume conservation is not  derived from a canonical formalism.}
in order to be able of regarding the  evolution operator $U$ as unitary, 
in the Hilbert space $ L^2 (R^{8n} , \eta )$. 
In other words one could perhaps go without a Hamiltonian structure, but not
 without a preserved volume element.
In practice we shall resort  to a  symplectic form anyway, in order to
get rid of the arbitrariness of the volume form.

\bigskip
\noi
{\bf 3.2 Density and distribution function}

\noi
Since velocities must point toward the future,  the relevant
 part of phase space is   the region of    $T(M^{4n}) $ defined 
  by    $  v \cdot v   >0,  \quad  v^0 >0 $.   
The "proper-time-dependent" density  
$ D (x_1,v_1, ....x_n,v_n, \tau_1, ....\tau _n) $
 is ruled by not one but $n$ Liouville equations
$$  ({\partial  \over \partial \tau _a}  +  X_a)  D  = 0      \eqno (3.5)  $$
This {\it  system\/} of equations was first written by Hakim for a gas of free 
particles.     R.Lapiedra and E.Santos  
pointed out that eq (3.3)   just ensures the integrability
conditions for $n$ Liouville equations in the presence of predictive 
interactions {\lapsan} (unfortunately their formulation is not thoroughly  
covariant, which leads them to restrict their statement to the first order in 
the coupling constants).

\noi  As emphasized in ref. {\hak}, only the "proper-time-independent" density 
$$ N = \int ^{+\infty} _ {-\infty}
D (x_1, v_1,.....x_n, v_n, \tau _1, ...\tau _n)
\  d\tau _1 .... d \tau _n         $$
has a direct physical meaning.
>From (3.5) it follows that  $N$ is a first integral, that is  $ X_a N =0 $.
Conversely, eqs (3.5) are formally solved by 
$ D = U_{\tau _1.....\tau _n }   N $,   provided $\lie _a N = 0$.

\medskip
\noi   When the masses are  specified from the start, the supports of $D$ and
$N$ are a priori restricted  to the region 
$ 2K_a = m_a ^ 2 $; owing to the condition that the velocities point to the 
future, one is left with a one-sheet mass shell.

\noi
{\sl Normalization and averages}  

\noi
The next step will be a manifestly covariant formulation for  phase averages 
and for  the  normalization of the distribution function.
We naturally use Hakim's definition {\hak}, that is  in our notation 
$ \int  N \alp  = 1   $
where  integration  is to be performed over  a $7n $ dimensional manifold 
cutting all the orbits of the evolution group.
{\sl Assuming that the motion generated by the $X_a$'s preserves some volume 
element $\eta$},  we are in a position 
to apply the results of section 2 in the dimension $m=8n$.   

\noi     Phase space is  $V^ {8n} = (T(M^ {4}))^ n  $ and  $X_1, ... X_n $
 are the generators of the motion. 
Now    $m-n = 7n $, and   $\eta $ induces  the $7n$ form 
 $$  d\Sigma _ {A_1...A_n} = {1 \over  (7n)!}  \
   \eta _{A_1,...A_n A_{n+1}...A_ {8n} }  \   
dy^ {A_{n+1}}  \wedge ...\wedge dy ^ {A_{8n}}      $$ 
  as   element of $7n$-dimensional hypersurface imbedded in $V^ {8n} $.

\medskip
 \noi
Since the Liouville operators are incompressible vector fields, we can apply 
  Proposition III of Section 2 and   conclude that  $ d(f \alp ) = 0 $ for 
any first integral $f$.    By the Stokes theorem,  $  \int f \alp $ over any
 {\it closed\/} $7n$-dimensional manifold  vanishes.
Notice that  in particular $f$ can be the distribution function.

\noi   We claim that the integral 
   $ \int f  \alp $  over a $7n$-dimensional surface  $S^ {7n}$
  does not depend on the choice of it provided it is specified  that $S^ {7n}$ 
cuts once and only once each orbit manifold (this precaution discards the 
possibility of integrating over the mass shell).
Since  we are dealing with a multitime flow,  this point 
is not so obvious as in the one-body case; the subsequent details are in order.

 \noi 
{\bf Definition} A {\sl flow tube } is an  $m$-dimensional {\sl invariant 
domain} of  $V^  {8n}$ (invariant under  the $n$-parameter group with 
 infinitesimal grenerators $X_a$).

\noi
Let $\cal T$ be a flow tube.      Our precise statement is that 

\noi  {\sl  The integral of $f \alp$ over the   cross-section of
 $\cal T$ by a $7n$-dimensional surface  $ S^ {7n} $ is independent 
of $S^ {7n}$, provided we assume that   $ S^ {7n} $ cuts (and only once) 
each integral curve of  $X_1....  X_n $}.

\noi  This can be proved as follows.

\noi     Let $\cal T$ be the orbit of a compact domain 
$ \Delta  \subset  S^ {7n}  $.
The cut of $ \cal T $ by another surface  $ {S'} ^ {7n}$ will be noted as
$ \Delta '$.
One must take some care of this complication (absent in the one-parameter 
case):   The fronteer  $\partial \cal T $ of the tube has  $8n-1 $ dimensions, 
whereas we need a $7n$-dimensional surface  in order to 
extend the set   $ \Delta   \cup  \Delta '$ as to form a closed manifold 
\footnote * {This complication was pointed out in ref {\hak}.}.

\noi  Fortunately it is clear that all the vector fields $X_a$ are tangent to 
  $ \partial \cal T$.
Thus $\alp $ vanishes on   $ \partial \cal T$.
Now it is   possible to construct a surface      $ {\cal B} ^ {7n} $
imbedded in        $ \partial \cal T$ as a submanifold, 
and connecting the domains $\Delta $ and $\Delta ' $ 
in such a way that     
$\Delta \cup \Delta ' \cup                {\cal B} ^ {7n} $
                       be a   closed $7n$-surface (see Appendix A).
Then we observe that $f \alp$ vanishes on     $ {\cal B} ^ {7n} $  for  it 
vanishes     on   $ \partial \cal T$.
Therefore applying the Stokes formula yields 
$ \displaystyle \int _{ \Delta '}  f \  \alp  =
 \int _ \Delta   f    \    \alp    $.

\medskip
\noi  Remark:

\noi  In applications we shall suppose that 
$S^ {7n} = S^ 3 _1 \times .....\times  S^ 3 _n  \times  (W^ 4) ^ n  $
where  $W^ 4$ is the space of four-vectors and  $S^ 3_a$ is a spacelike 
three-surface imbedded in Minkowski space.

\noi
Besides the normalization of $N$, 
another application concerns the total {\sl entropy } 
  $ S = - \int   N  \log N  \alp $. 
Since the phase function $N \log N $ is a constant 
of the motion, under similar assumptions  $ N \log N \alp $ is a closed form,
 a result  which renders   $\int  N \log N  \alp  $ independent of 
the integration surface.

\noi
More generally, we have an intrinsic   covariant definition  for 
the average of any constant of the motion.

\medskip
\noi
In general  a form like $\eta $ is by no means unique.
However it may happen, for a given system of $n$ interacting particles, 
that     the $8n$-form  $ \sum  d^{4n} x \wedge d^{4n} v $ is 
"accidentally" conserved by the motion, providing obviously a preferred choice
(see Appendix B for an example at first order in the coupling constants).

\noi  Otherwize there remain the possibility of redefining the distribution 
function (without changing parametrization) whenever we change the preserved 
volume form. So doing we keep $N \alp $,  and therefore the formulas for 
averages,  unchanged.

\noi  Most simply, we shall resort to the Hamiltonian formalism (next section).
Although a similar ambiguity arises in the Hamiltonian framework, it can be 
removed by an  argument which is plausible for all gases made of microscopic 
"molecules".

\bigskip

\noi
{\bf  Reduction to 6n dimensions}.

\noi
For free particles it was already pointed out by Hakim that 
normalizing over a 
$7n$-dimensional surface actually reduces to a customary normalization over 
a  $6n$-dimensional surface, by help of the $n$ mass-shell constraints 
 $ m^ 2   =  v^ 2 $.
It is fortunate that the multitime formalism of relativistic dynamics still 
provides  $n$ mass-shell constraints in the presence of interactions.
These constraints amount to fix $n$ constants of the motion 
$ K_1, ....K_n $  identified as $ {1 \over 2} m_a ^ 2 $.

\noi 
In most cases of interest, the relevant distribution functions will be 
 concentrated on a  submanifold 
$  2K_a = m_a ^ 2 $, with specified values of $m_1, .....m_n$.
Therefore we expect that the distribution function (as defined in the whole 
phase space) involves a  factor $ \prod  \delta (m_a ^ 2 - 2K_a) $.

\noi
Since  the velocities  must point to the future, there is    in  $N$
 a  factor 
$  \prod  \theta (v_a ^ 0 ) $, where $ \theta $ is the step function.
These considerations   entail that the effective phase space may be 
 $6n$-dimensional, as expected for a  contact with the newtonian limit.

\noi  Remark:

\noi  Restriction to a sharp mass shell agrees with the realistic picture of a 
gas of particles with given masses; but it is  not imposed by our formalism.
Moreover,   a statistics of particles with unspecified 
masses may be of interest for cosmological applications and for hadronic 
matter.
In addition a smearing of the mass shell becomes  technically necessary when 
entropy is defined as above, for the logarithm of a Dirac distribution has no 
mathematical meaning.

 \noi
In fact the essential  mass-shell property is   that the quantities 
$  m_a ^ 2 - 2K_a  $    are constants of the motion. But strictly speaking they 
are not constraints untill one gives  numerical values   to these constants,
 as happens when we assume the factors  $\delta (m ^2 _a  - 2K_a )$.

\medskip
\noi
We shall see in Section 4 that the computational 
complications introduced by the factors 
considered above disappear  in the special case of " perfect interactions".

\bigskip
\noi
 We could hardly go much further without a Hamiltonian formalism; indeed:
as soon as one whishes to consider either energy or total linear momentum, a 
symplectic canonical formulation is badly needed.

\bigskip
\noi
{\bf 4. Hamiltonian formalism}

\noi
The hamiltonian formulation consists in looking for a symplectic form 
$ \Omega $ invariant under the evolution group, that is 
 $ \displaystyle  \lie _{X_a} \Omega = 0 $ for all $a$.
 When an invariant symplectic form $ \Omega $ has been determined, the 
Liouville operators are generated by  positive scalars  functions 
$ H_a $ satisfying  $ i_a \Omega = d H_a $. We refer to these scalars as the 
{\sl Hamiltonians},
 although they are associated to (half) squared masses instead of energy 
{\dvlxxv}.
In any system of canonical coordinates $q_a ^\mu  , p_b ^\nu $, we can write
 $  \Omega = \sum dq_a ^ \alp  \wedge dp_{a \alp} $. 

\noi       The volume form is
    $ \eta = d^ {4n} q \wedge   d^ {4n} p $. 
Notice that      $\eta = const. \   \Omega ^ {4n} $ (exterior power).
Since $\lie _a \Omega = 0 $, the Liouville theorem follows 
$  \lie _a  \eta = 0 $. 

\noi 
   The scalar density $\gamma $ is equal to unity in canonical 
 coordinates; this is not always the case in natural coordinates $x, v$.  
  In arbitrary  coordinates, the volume element can be written as
$\eta =  \gamma  \    d^ {8n} y  $ where 
$\displaystyle    \gamma  =  {D(q,p) \over D(y)}  $, 
with an obvious notation for the Jacobian.
 Drastic simplification arise when the position formulas "accidentally" imply 
that 
$d^ {4n}  x \   d^ {4n} p = d^ {4n}  q \   d^ {4n} p $.

\bigskip
\noi
In contrast to Newtonian mechanics, and in view of a famous theorem {\cur},
 it is not possible to require that the physical positions be canonical 
variables.       In other words  we cannot have just $q_a = x_a $ 
throughout phase space. This situation implies that, for a given system of 
Liouville operators $X_a$, the invariant symplectic form $\Omega $ cannot be
 unique.
 This complication is a source of  ambiguities  
not only about the Hamiltonian formulation but also in the definition of
 linear momentum   (but, insofar as one is concerned only with finding first 
integrals and with solving equations of motion, there  is  in principle 
nothing wrong with the fact that a given physical system  admits infinitely 
many symplectic formulations).

\noi
We notice that  quantization is naturally affected by this  ambiguity, but
 this is not at  all a
 difficulty  because there is no reason why a classical system should 
correspond to a unique quantum system; we expect  unicity when going from 
quantum to classical theory and not  the reverse! 

\noi
Remaining in the classical area of physics, the situation becomes more 
serious  as soon as we try to construct statistical mechanics with concrete 
applications in mind.

\noi
If statistical mechanics were limited to formal manipulations involving some 
preserved volume element of phase space, again there would be no problem in 
having, for a given physical system of many particles, infinitely many 
possible descriptions.
But we are interested in the construction of macroscopic quantities like the 
thermodynamical functions.
At this stage the Hamiltonian formulation seems to play a crucial role.  
Several basic concepts like the energy or the canonical linear momentum are 
intimately related with this formulation through Noether's theorem,
 and it seems that the definition of
 most macroscopic quantities   necessarily involves these concepts; 
 temperature is the first  example. 
Finally the very notion of 
equilibrium   rests  on the previous determination of a canonical formalism
\footnote * 
{One might speculate that there exists some unifying formula defining 
 temperature, entropy and  equilibrium, in a way which is invariant 
under reparametrization and changes of symplectic form.
But for the moment nobody has any idea of such a formula, if it exists at 
all;   therefore one has to  cope with the ambiguities.}.

\medskip
\noi
The lack of unicity in the symplectic formulation can be expressed in saying 
that  the dynamics of a relativistic system {\it cannot\/} be completely 
determined by the simple knowlege of the Hamiltonians.

\noi  However it is true that, to a large extent, spectral and asymptotic 
 properties of relativistic dynamical systems are 
fully  determined by the Hamiltonian generators of the motion.
For example we have been able to speak of the "abstract integration" of
 a system {\dvlxxv}.
 Abstract integration refers to the solving of the canonical 
equations of motion, in terms of $q,p$,  and 
provides a foliation of phase space by $n$-dimensional leaves, disregarding 
the way how each leaf is identified as a cartesian products of $n$ 
 lines  which are the lifts of world lines.
In the same spirit, but in the language of constraints theory, 
it was observed by Todorov \constr that, under very general assumptions,
 the scattering properties of an $n$-body 
system depend only on the form of the mass-shell constraints, and {\it not\/} on 
the gauge fixations.

\noi
In other words we can see a relativistic  hamiltonian  system as an 
{\it abstract structure\/} (corresponding to the $H_a $ only) 
completed by  additional  formulas 
 connecting the physical positions  $x^\alp _a $ with the canonical 
variables. These {\it position formulas\/} must satisfy the {\it position 
equations\/} \dvlxxv
$$  \{   x^\alp _a ,  H_b    \} = 0  \quad \forall a \neq b 
                                                            \eqno  (4.1)  $$
necessary for achieving the determination of world lines.

\bigskip
\noi
To summarize, the ambiguities of   the Hamiltonian formalism are of  two kinds.

\noi
1)Any change of parametrization implies a redefinition of the "velocities" and 
a redefinition of the set of vector fields $X_a$.

\noi
2) Even when the parametrization is kept fixed,
 it is in general impossible to find 
a {\it unique\/} symplectic form invariant by these vector fields.

\noi
These ambiguities are the  price paid for a  covariant setting.
They cannot arise  in Newtonian  mechanics for two reasons:
the use of an absolute time and the implicit prescription that the physical 
positions are canonical variables.
But in relativistic dynamics, considering mutual interactions forbids to
 require that the physical positions $x_a $ be canonical variable.

\bigskip
\noi
At this stage we are led to the following observation.
Statistical mechanics is a very general theory; its applications may concern 
a fluid of galaxies as well as an ordinary gas made of diatomic molecules.
We suggest to distinguish clearly these two situations. 
For example the concept of 
temperature has certainly a profound physical significance when it concerns the
 air   we breath.
When applied to a gas of which the "molecules" are galaxies, this concept has 
hardly the same physical meaning, although it is defined through the same 
mathematical structure.
Therefore we can provisionnally consider that some ambiguity about the 
statistical mechanics of a very large number of {\it  macroscopic bodies\/}
 is perhaps not so scandalous after all. Naturally further work is needed in 
order to select a convincing prescription removing the ambiguity in this case.

\noi
In contradistinction the description of a gas made of {\it microscopic 
molecules\/} should not suffer from the same uncertainty.

\noi
Fortunately,  at the scale of an ordinary gas it is reasonable to remember 
that a  classical system of $n$ microscopic particles must be considered as
 the classical limit of a quantum system.
Our point is that if we carefully formulate the axioms of $n$-body quantum 
mechanics, then taking the classical limit of a system  will
 automatically determine a particular Hamiltonian formulation.
In order to prove this, let us first sketch the framework for quantum 
mechanics of $n$ interacting particles.
The wave equations involve $n$ half-squared-mass operators 
$ \displaystyle   (H_1)_{op} ,   .....(H_n) _{op}     $
acting in some Hilbert space and commuting among themselves \dvqu
(this point of view stems from quantization of predictive mechanics, but it 
also agrees with constraint theory).
However the  knowledge of these operators is not sufficient for a complete 
determination of the system.
This point is widely overlooked in the litterature, because most investigations 
are concerned with either spectral or scattering properties fully contained 
in the squared-mass operators,
 all question  about position measurement being systematically ignored.

\medskip
\noi
Our proposal was that there exist  $n$ {\sl coordinate operators}
  with four  components, say
    $ \displaystyle  (x^ \alp _a)_{op} $
satisfying the commutation relations 
$$ [       (x_a ^ \alp)_{op}  ,  (H_b )_{op}    ]   = 0     
    \qquad  a \neq  b        \eqno (4.2)  $$
For instance these $(x)_{op}$'s are relevant objects   if one whishes  to 
   consider position measurements 
\footnote * {These "coordinate operators cannot,
as they stand, be considered as position operators in the usual sense. They 
   act on some subset of $L^ 2 (R^ {4n},  d^ {4n}x ) $, 
which is not the case of the Newton-Wigner operator.}.

\noi
If the system has a classical limit, these relations reduce for 
 $\hbar \> 0 $ to the 
   "position equations"  (4.1) of predictive mechanics {\dvlxxv},
 and the coordinate operators  $x_{op} $ are expected to have a unique and 
well-defined limit 
$ x_a  ^ \alp (q,p) $. 
>From this  limit we can derive the generalized Legendre transformation 
$  q,p  \longleftrightarrow  x,v $,
 which also  determines the rescaling from  the parameters
$ \tau _1, ....\tau _n $ to the proper times. 

\noi
To summarize, the classical limit is expected to determine a predictive 
dynamical system {\it endowed with a preferred invariant symplectic form\/} 
(corresponding to a preferred  Hamiltonian formulation).

 \noi   This argument solves an important question of principle.
But in practice, the "good" symplectic form will remain ignored as long as the 
details of the underlying quantum theory remain ignored.
This might be the case of "predictive electrodynamics" \belfus untill it will 
be   rederived  from some $n$-particle generalization of the Bethe-Salpeter 
equations.


 \bigskip
\noi
{\bf 4.1. Perfect interactions}

\noi
Ideal gases are made of noninteracting particles.
The next step consists in considering an interaction which can be trivialized 
by a suitable  transformation.
With this idea in mind let us characterize a {\sl perfect interaction}  by 
the  global existence 
of particular canonical coordinates, say  $\qat, \pat $, allowing to write 
$$ H_a = {1 \over 2} \pat ^ 2 _a      \eqno (4.3) $$
We refer to  $ \qat , \pat $ as  {\sl Hamilton-Jacobi (HJ) coordinates} and 
we shall say that we have an {\sl almost ideal gas}.
Let us stress the following point:
whereas (4.3) could be {\it locally\/}
 written    for any Hamiltonian  system by 
solving an $n$-body relativistic generalization of the Hamilton-Jacobi 
equation,  the existence of {\it global coordinates\/} like $\qat, \pat $ is a 
very particular property  of the gas we consider {\hj}.

\noi
A system undergoing perfect interactions   is spectrally trivial; 
the Lie algebra of first integrals has the same structure as in the free 
system. In  principle equations of motion  can be exactly solved.
Nevertheless,  word-lines actually deviate from straight lines.   
It is clear that perfect interactions realize the most simple situation 
beyond the case of  noninteracting particles.

\bigskip
\noi
We strongly suspect that
 considering HJ coordinates  amounts to    neglect bound states.
This should not prevent us from using perfect interactions as a 
simplification when attractive forces are present.
This approximation  may be applied to the case of attractive forces, 
provided that  the gas we  consider is so   {\sl dilute } that it 
 is reasonable to neglect bound states.
Naturally one must keep in mind that this representation is valid only 
insofar as the distances between particles are large enough.

\medskip
\noi
The main interest of HJ coordinates 
lies in the possibility of explicitly writting down covariant many-body
 interactions.   Indeed, 
for an almost ideal gas, the most  general interaction  can be explicitly 
written  in terms of the HJ coordinates.

\noi  According to (4.1), the components of  $x_a $ must be     invariant by  
all the vector fields $X_b $ generated by hamiltonians $H_b$ where $b\not= a$.
We now have to solve
$ \displaystyle   \{ x_a,  \widehat p ^2_b \} =0 $. We immediately find
                         $$ x^ \alpha _a =
\widehat q ^ \alpha _a   +   F_a^ \alpha
 (....\widehat p_b,  \   \widehat  q_b  \wedge \widehat p_b ...   ) 
 \eqno (4.4) $$
with all $b \not= a $ in $F$, and $F$ are arbitrary vector-valued functions
 except for the following restrictions:

i) They might be required to satisfy reasonable boundary conditions,
namely that  in some sense $ \qat , \pat  \>  x , v $  at temporal or spatial 
infinity.

 ii) They must  respect  Poincar\'e invariance.
Scalars invariants and   {\sl vector  invariants}  of the Poincar\'e group 
can be characterized as well in terms of HJ 
coordinates. According to this remark, $F^ \alpha _a $ in the position formula 
above must be a vector invariant. 

\noi
Of course,  the simplest solution    $x_a  = \widehat q_a  $
 would be trivial;      it corresponds to straight world lines.

\medskip
\noi It might be interesting to discuss under which conditions  the 
 Feynman-Wheeler electrodynamics can be 
(approximately) treated as a perfect interaction.

\bigskip 
  \noi
Using HJ  coordinates  we can write 
$X_a =   \pat _a ^\alp  \    \partial /   \partial \qat _a ^\alp  $.
For the special choice of a surface $S^ {7n} $ defined by  equations of the 
form
$$ s( \qat _1) =    s( \qat _2) = ......      s( \qat _n) = 0   $$
(for instance  $s(q) = q^ 0$)  we   have this simplification   for the 
$7n$-form $\alp$ of Sections 2,3 
$$ \alp = const.  \         \pat _1 ^{\mu _1} ..... \pat _n ^{\mu _n}     \
d \sigma _{\mu_1} (1) \   \wedge ..... \wedge  d\sigma _{\mu_n}(n) \
\wedge   d^{4n} \pat        +  O(S^{7n} )    \eqno (4.5)   $$
where  $ O(S^ {7n}) $  is a term  vanishing  on  $S^  {7n} $ 
and this  notation
$$ d \sigma _ \mu (a) =
{1 \over  3! } \varepsilon _ {\mu \nu \rho \sigma }  \  
 d \qat ^\nu _a  \wedge   d \qat ^\rho  _a    \wedge  d \qat ^\sigma  _a   $$
$$ d^{4n} \pat = d  ^4 \pat _1 \wedge   ..... \wedge  d ^4 \pat _n   $$
Then with help of the identity 
$ d ( p^ 2)  \wedge  d^ 3 p  \equiv 2 p^ 0 \   d^ 4 p $ 
we factorize $ d ( \pat ^ 2  _a ) $  in  $ d^ {4n}  \pat $ and finally,
for all "phase function"  $\Phi $ we can write
$$    \int _{S^ {7n}} \Phi \alp =
const.    \int _{S^ {7n}} \Phi  \   d( \pat ^ 2 _1 ) \wedge ....
\wedge d (\pat ^ 2 _n)  \wedge  \lambda    \eqno (4.6)  $$
where $\lambda _{ (X, \eta)}$ is a  $6n$-form which also depends on the choice 
made for $ S^ {7n}$.
When  $\Phi $  has the particular form 
$ \Phi = \theta (\pat ^ 0 _1) ....\theta (\pat ^ 0 _n)
 \delta (\pat ^ 2 _a - m^ 2 _a ).....    \delta (\pat ^ 2 _n - m^ 2 _n )
\widetilde \Phi   $,
then equation (4.6) is reduced to  
 $$          \int _{S^ {7n}} \  \Phi \   \alp = 
const. \int _ {S^ {6n}}  \widetilde \Phi  \  \lambda   \eqno (4.7) $$
where  $  S^ {6n}  =   S^ {7n}   \cap  mass\  shell $.
This formula makes the contact with more popular formulations using a 
$6n$-dimensional phase space; also it is an illustration of a remark  made
 in section 3.2.

\bigskip
\noi    It is clear  that all first integrals of the interacting system
 are known in terms of HJ coordinates, 
 which allows for  {\it formally\/} solving the Liouville equations (3.5).
But the solutions obtained in this way are expressed in terms of canonical 
coordinates.  For practical purpose it is usually of interest to have 
solutions in terms of the natural coordinates $x,v$, which in turn requires 
knowledge of the position formulas.

\bigskip
\noi
{\bf 4.2.  Average  momentum } 

\noi
In the framework of Hamiltonian formalism, in the absence of external forces 
the system is invariant under translations  and rotations, hence the first 
integrals  $P = \sum p_a ,  \quad  M = \sum  q_a  \wedge p_a  $. Application 
of Proposition III yields an intrinsic  definition of   the averages
 $  \displaystyle 
\overline P^ \alp, \quad  \overline M ^ {\mu \nu} $. For instance 
$$ \overline  P  ^ \mu =
\int N P^ \mu   X_1 ^{ A_1} ...X^ {A_n} _n  \  d\Sigma _ {A_1...A_n}  $$
 is conserved under a change of the   $7n-$dimensional integration manifold.

\bigskip
\noi   {\bf 5.  Canonical Distribution. Equilibrium}.

\noi
Provided the system is translation invariant, it is natural to define 
equilibrium by the distribution function
$$  N = const. \   
 \delta ^ +_1  .....\delta ^ +_n  \ 
   \exp (- \beta ^ \mu  P_ \mu)    \eqno (5.1) $$
    for some constant timelike four-vector $\beta ^ \mu$ associated with the 
inverse temperature. The factors 
  $$  \delta ^ + _a  =   2  \theta ( v_a ^ 0)  \   
  \delta (m_a^ 2 - 2H_a)          \eqno (5.2)  $$
take into account the mass-shell constraints and the direction of the arrow of 
time on world lines.

\noi
 For an ideal gas, equation (5.1)
 (which is also an $n$-body generalization of   J\"uttner's formula)
agrees with the definition proposed by Hakim three decades ago 
\footnote *{Formula (5.1) seems to differ  from eq (6.15) of  \hak  by the 
factors $\delta ^+$. These factors are implicit in ref. \hak  
where the reduction of phase space by mass shell constraints is assumed from 
the outset.}. 

\noi
In the presence of  interactions, $v_a ^0 $ is not constant, but 
   $\theta (v_a ^ 0)$ is a (discrete)
first integral although $v_a ^ 0 $ is not.
Finally  $N$ is a first integral as it should.

\medskip
\noi
Let us stress that there is in general no indication that the replacement of 
  $\theta ( v ^ 0 _a) $  by $ \theta (p_a ^ 0) $ be legitimate.
 Strictly speaking, in order to write down the equilibrium distribution 
function in closed form, 
one might be obliged to solve the position equations! 
This is no surprise; it is reasonable that some information from world lines 
be necessary.
However, interesting and drastical simplifications arise in some special 
cases, as will be briefly discussed below.

\noi
Being spatially homogeneous, the equilibrium distribution function is not 
normalizable, which rises the problem of elaborating a covariant procedure for 
the thermodynamical limit.

\medskip
\noi
In the presence of an external field, space translation invariance is broken, 
whereas it often happens that the energy  
 $   l ^\mu  P _\mu   $ remains conserved 
(the  external potentials  applied to the system must be stationary). 
In this case     $\{ l^ \mu  P _ \mu , H_a \} = 0   $
 for some constant timelike {\it unit vector\/}  $l$ 
defining a "laboratory frame".                      Then we define 
equilibrium by 
$$ N = const. \  \delta ^+_1 ....\delta ^+ _n \  
  \exp (-  \beta  \     l^\mu P_ \mu)        \eqno (5.3)   $$
with  $\beta$ a positive scalar.
 Again, one is led to consider the thermodynamical limit. But the analogy with 
the Newtonian situation is more transparent here.
However, existance of the limit is an open problem. For instance it was 
argued that for long-range interactions, it may happen that this limit does 
not make sense {\lapbar}).

\bigskip
\noi
Returning to  translation invariant systems, it is clear that the natural 
coordinates  $x,v$ allow for a simple expression of the factors $\delta ^ +$ 
(at least in the autochronous case); but $\gamma $ may still differ from 
unity in these coordinates.

\noi
In contrast, 
any choice of canonical coordonates provides a trivial  expression for  $P$, 
but this simplicity is at once destroyed when taking into account the
 mass-shell constraints.
  In any frame adapted to $\beta$ we get 
$ N= const. \  (\prod \delta ^ + )   \   \exp (- \beta  P ^ 0) $,
 where $P^ 0$ is 
the sum of the $p_a^ 0$'s.
 In ordinary canonical coordinates  $p^0 _a $ takes on the form 
  $ \displaystyle 
p_a ^ 0 = \sqrt {m_a^ 2 + {\bf p}_a^ 2  - 2 V_a (q,p) }  $.
Also the  factors $\delta ^ +$ are in  general  model dependent, since the
 canonical expression of $H_a $ in terms of  $q,p$ can be complicated.

\medskip
\noi But in the special case where  HJ-coordinates actually exist,  remarkable 
simplifications arise.  

\noi First, the mass-shell constraints take on a very simple form (in fact as 
simple as for free particles).
  In HJ-coordinates 
 $ \widehat  p _a ^ 0 = \sqrt {m_a^ 2 +  \widehat {\bf p}_a^ 2 }  $.

\noi    Second,  
we can legitimate the replacement of $v^ 0 $ by $p^ 0 $ in the step functions.
Indeed,  $H = {1 \over 2} \widehat p ^2 $ is a constant of motion, 
it is clear that on each orbit $\widehat p$ remains timelike. Since $p $ 
reduces to $ v $ 
for a vanishing coupling constant, it is clear that $ \widehat p $ as well as 
$v$ is oriented toward the future.
In other words, $\widehat p$ and $v$ are simultaneously in the future light 
cone.     Third,  $ P = \sum  \pat $.

\noi
Finally the equilibrium distribution takes on the free form when expressed in 
terms of HJ coordinates.

\bigskip
\noi
{\bf Microcanonical equilibrium}

\noi Let us consider a gas of interacting particles in the absence of external 
forces. We suggest defining the microcanonical ensemble by a  straightforward
 generalization of eq (A4) of ref {\hak}, say
$$ N = const. \  
\delta (\zeta ^ \mu - P ^ \mu) \    
\delta ^ + _1 .....\delta ^ + _n      \eqno (5.4)     $$
where $ \zeta $ is a constant timelike vector and  $\delta ^ + _a$
 is given by   (5.2).
Again, the above distribution is a constant of the motion.

\bigskip
\noi
{\bf 6. Conclusion}

\noi   The theory of direct interactions  provides a solid ground for a  
covariant formulation of  statistical mechanics.  Let us be more specific: 
predictive relativistic mechanics is the most natural way for taking 
seriously the idea (already present in Hakim's early work) that one is 
dealing with an  $n$-parameter evolution group, associated with a multitime 
parametrization.             As we know for a long time,  
this philosophy can be expressed in terms of   modern differential geometry,
which  leads to  vector fields in $8n$-dimensional phase space; these 
"Liouville operators" have  mutually vanishing Lie brackets, 
ensuring that initial velocities and positions actually determine world 
lines.

\noi
If we now consider a statistical mechanics of relativistic particles,
 the predictivity conditions are  technically essential; 
 on the one hand they entail  integrability of the $n$ 
 Liouville equations (it is an exact result, by no means 
limited to first order in the coupling); on the other hand they ensure, as 
we have shown,   the intrinsic nature of important 
  quantities defined by  a flux integral.
For instance they imply the conservation of phase average values, especially  
the conservation of norm (numerical flux of particles) and entropy.   A few 
manipulations of exterior differential algebra  were sufficient to demonstrate
 that  
predictive mechanics  exactly fulfills the demands of relativistic statistical 
mechanics concerning the Lorentz invariance of the phase averages. 

\noi
In our  exposition we have as much as possible separated the specific 
features based upon  the Hamiltonian formalism   from more general topics
 requiring simply a  preserved volume element.

\noi   Insofar as one aims at a general theory devoted to  a description 
of systems out of 
equilibrium, the basic notion is that of a preserved volume element, rather 
than a symplectic structure:  in other words the Liouville theorem without the 
Hamiltonian formalism!
This viewpoint is close to the  spirit of   mathematical ergodic theory.
However we notice that, for a system defined by its equations of motion, the 
preserved volume element is not univocally determined.

\noi
In this work we have focused on the $n$-particle distribution function. But we 
expect that also {\it reduced\/} distribution functions can be proved to satisfy 
(in the presence of mutual interactions) covariant evolution formulas and will 
enjoy  conservation properties with respect to normalization.
This point is obviously related with the important issue of a tractable  and 
relativistic BBGKY hierarchy.

\noi
A Hamiltonian formulation is highly desirable anyway, as soon as one is 
interested in a truly physical description.
In the presence of mutual interactions, this formalism  is essential in order 
to define linear momentum and energy.
Thus   the notion of  equilibrium and probably also  the construction of 
thermodynamical functions require a  Hamiltonian formulation.

\noi
At least for gases made of microscopic constituents, a certain idea of the 
underlying quantum mechanics solves (in principle) the ambiguity related with 
the impossibility of having canonical physical positions.

\medskip
\noi
The usual expression for the  equilibrium distribution function has been 
generalized in a covariant formula  taking mutual interactions into account.
Further work is needed in order to check in which sense and in the context of 
 which side  conditions this definition actually 
realizes the extremum of the entropy integral.

\noi
There is a good hope that several methods and formulas concerning ideal gases 
 remain valid when the 
mutual interactions are modeled as "perfect interactions".
This simplified picture provides a toy-model of interacting particles for 
which rigorous results can be proved.
Its relevance for a real gas is probably  limited to the dilute-gas 
approximation, but many  standard calculations of the theory of 
ideal gases can be extended to  almost ideal gases in a straightforward fashion.

\medskip                         
 \noi   Rather than new practical applications,  the scheme proposed here 
opens theoretical possibilities.  For instance it sheds light on the 
transformation properties of various  results available
 untill now in   noncovariant form.

\noi                                          
The present formulation allows for considering particles with unspecified 
masses, which is interesting for applications to cosmology or to hadronic 
matter (the latter suggests an extension toward {\it quantum\/} relativistic 
statistical mechanics).

\noi Several generalizations are possible; for example one could enlarge phase 
space as to include radiation.

 \noi
As it stands our exposition  may seem to be 
exceedingly geometrical, but this feature is an advantage in view of possible 
generalization to curved spacetime.

\noi
A comparison with the "statistics of events" developed by Horwitz {\it 
et al.\/} \horstat will be of interest.

  \bigskip
  \centerline {\bf Appendix A}

\noi
Through any $y \in \Delta $ passes a unique orbit, say $\Gamma _y $ of the 
evolution group.
According to the assumptions made, this orbit is cut in a unique point $y'$ by 
the surface  $ {S'} ^ {7n} $. We can write 
$$ y' = U_{  {\tau '}_1  {\tau '} _2 ....{\tau '} _ n}  \  y  $$
for a sequence of parameters 
 $  \displaystyle    {\tau '}_1 ,  {\tau '} _2 ,.....{\tau '} _ n     $ which 
depends only on $y$ and on the cutting surface. 
Among all possible curves lying in $\Gamma _y $, a very simple way for joining 
$y$ to $y'$ is characterized for instance  by       the conditions
  $$  { \tau _a  \over \tau _1 }  =
   { {\tau '} _ a    \over   {\tau '}_1 }                $$ 
Indeed $\Gamma $ is endowed with a network made of the integral curves of 
$ X_1 ,...X_n $.
This procedure defines a distinguished curve $C_y $ which lies in  $\Gamma _y$ 
and joins $y$ with $y'$.

\noi
Now when $y$ runs in the $(7n -1)$-dimensional  fronteer  
 $ \partial  \Delta $ of the domain $\Delta$, the curves  $C_y$ generate a
$7n$-dimensional    surface  $ {\cal B} ^ {7n} $ connecting 
            $ \partial  \Delta $   with        $ \partial  \Delta ' $.
Notice that two different $\Gamma $ cannot intersect unless they are 
identical.            
When $y$ runs in      $ \partial  \Delta $ the $n$-dimensional orbits 
$\Gamma $ generate  $ \partial {\cal T} $.
Since each $C_y $ lies in $\Gamma _y $ it is clear that, in the generic case,
$  {\cal B}^ {7n}   \subset \partial  {\cal T} $ as a submanifold.

\noi  
In this construction, $   {\cal B}^ {7n} $
 can be considered as the fronteer of a
"sub-tube" generated by the integral curves of a vector field  
$ \Xi  = X_1 +  \nu_2 X_2 + ..... + \nu _n X _n     $  where  the $n-1$ 
independent   quantities $\nu _2 , ....\nu _n $ are    first integral of the 
system    $X_1, ....X_n $.

  \bigskip
  \centerline {\bf Appendix B}

\noi  {\sl       At first order   in the coupling constant,
the $8n$-form $\eta \zer = d^{4n} x  \wedge d^{4n} v  $ is preserved by
 predictive electromagnetism and by predictive vector interaction}.

\noi  Proof

\noi  At this  order of approximation, the acceleration on particle $a$ is 
given by  \belfus
$$ \xi _a ^\alp = 
\sum _ {b \neq a} F^{\alp \sigma} _ {a \< b}   \  v_a ^\sigma   $$
where  the skew-symmetric tensor  $  F_{a \< b} $,
is (up to a constant factor) the Lienart-Wiechert 
field created at point $x_a$ by the  (vectorial or)  electric charge  
with label $b$ supposed to move on a straight line in spacetime.
Therefore, $ F _{a \< b} $ depends on the $v_b$'s  for   $b \neq a$, but 
certainly not on  $v_a$. Hence
$  {\partial  \xi ^\alp _a    /   \partial v_a ^\sigma} =
 F ^\alp   _ \sigma      $.                                  
By contraction of indices  
$   {\partial  \xi ^\sigma  _a   /   \partial v_a ^\sigma} = 0       $.
Since we are using the natural coordonates  $x, v$,  this equation amount to 
write     
$  \partial _ A   X ^A = 0 $, which proves our statement.

\bigskip
The author is grateful to Prof.R\'emi Hakim for a stimulating discussion.
  
\bigskip
 \item {\lich} A start to covariant kinetic theory was given by
  A.LICHNEROWICZ, R.MARRIOT,  
 Compt.Rend.Acad.Sc. Paris, {\bf 210},  759 (1940).

\item {\hak} R.HAKIM,    Journ. Math. Phys. {\bf 8}, 1315 (1967).

 \item {\isrkan} W.ISRAEL and H.E.KANDRUP,
 Ann. of Phys. (N.Y.) {\bf 152}, 30-84  (1984).
These authors have explicitly stated that the 
gravitational forces at work between "particles" should be modeled by direct 
interactions, and they cared to consider $n$-dimensional world sheets in an
 $8n$-dimensional phase space.

\item {\fw} J.A.WHEELER, R.P.FEYNMAN, Rev. Mod. Phys. {\bf 17}, 157 (1945).

 \item {\vdw} H.VAN DAM and E.P.WIGNER, Phys.Rev. {\bf 138 B}, 1576 (1965).  

 \item {\aadis} {\sl Relativistic Action at a Distance, 
Classical and Quantum aspects}, Edited by J.Llosa, Lecture Notes in Physics 
162, Springer-Verlag, (1982).

 \item {\dvlxx}  Ph.DROZ-VINCENT, Lett.Nuov.Cim {bf 1}, 839 (1969); 
Physica Scripta, {\bf 2}, 129 (1970).

\item { \dvlxxv}     Ph.DROZ-VINCENT, Reports in Math.Phys.  {\bf 8}, 79 
(1975); Ann.Inst.H.Poincar\'e {\bf A 32}, 317 (1980).

\item {\hakII} This scheme could be extended to second order derivatives in 
order to take radiation into account along the line of 
R.HAKIM,  Journ. Math. Phys. {\bf 8}, 1379 (1967).

\item {\belpredic}   Along the line of R.N.HILL and E.H.KERNER, 
Phys.Rev.Letters {\bf17}, 1156 (1966),
 L.BEL has proposed an alternative version of  predictive mechanics which is 
not manifestly covariant (although relativistically invariant) 
and will not be discussed here: Ann.Inst.Henri 
Poincar\'e, {\bf A 12},307 (1970); {\it ibid.\/} {\bf A 14}, 189 (1971).
In contrast, H.P.KUNZLE has considered a  generalization of the 
covariant formalism: Journ.Math.Phys. {\bf 15}1033 (1974). This extension is 
tremendously  complicated for our purpose.
           
\item {\lapsan} R.LAPIEDRA and E.SANTOS,  
  Phys. Rev. D {\bf 23}, 2181-2188 (1981).

\item {\horpir} L.P.HORWITZ, C.PIRON, Helv.Phys.Acta  {\bf 46}, 316 (1973).

\item {\horstat}  L.P.HORWITZ, W.C.SCHIEVE, C.PIRON, Ann.Phys. {\bf 137}, 306 
(1981).
L.P.HORWITZ, S.SHASHOUA, W.C.SCHIEVE, Physica A {\bf 161}, 300 (1989).
L.BURAKOVSKY, L.P.HORWITZ, Physica A {\bf 201}, 666 (1993).  

\item {\berg}  P.G.BERGMANN, Phys.Rev. {\bf 84}, 10266 (1951). 

    \item  {\abr}   R.ABRAHAM, MARSDEN, {\it Foundations of 
Mechanics, 2nd Edition\/}, Chap.2, exercise 2.4B, p.121, Addison,Wesley (1978).

\item   {\aren}  For the two-body case see    for instance
  R.ARENS, Nuov. Cim. {bf B 21}, 395, (1974).                   
  Here,  the formula  with two wedges is a short-hand notation
 for a pseudo-vector in space-time.

    \item {\dvlxxvi}    Ph.DROZ-VINCENT,  Contribution 
to {\it Differential Geometry  and Relativity\/}, Edited by M.Cahen and M.Flato,
 Reidel Publishing  Dordrecht-Holland (1976).

   \item {\dvlxxvii}        Ph.DROZ-VINCENT, Ann.Inst.Henri Poincar\'e,  {\bf A 
27},407 (1977).

\item {\lus}  L.LUSANNA, Nuovo Cim. A {\bf 64}, 65-88 (1981).

 \item {\constr}  I.T.TODOROV, JINR Report E2-10125, Dubna (1976); see also
contribution to {\aadis}.
V.V.MOLOTKOV, I.T.TODOROV, Comm.Math.Phys. {\bf 79}, 111 (1981).

 \item {\byac} U.BEN-YA'ACOV,
  Irreversibility in Relativistic Statistical Mechanics,
 Modern Physics Letters B, {\bf 8}, 1847-1860  (1994).

\item {\hunz} W.HUNZIKER, Comm.Math.Phys. {\bf 8}, 282 (1968).
        This author assumes a Hamiltonian formalism; this technical assumption
is  not conceptually required, though it seems to be necessary in order to 
derive the results.

 \item {\dvscat}         Ph.DROZ-VINCENT, Nuov.Cim. {\bf A 58}, 365 (1980).

\item   {\horrohr}  L.P.HORWITZ, F.ROHRLICH, Phys.Rev. D {\bf 24}, 1528 (1981).

\item {\cur} D.G.CURRIE,
 T.F.JORDAN, E.C.G.SUDARSHAN, Rev. Mod. Phys. {\bf 35}, 350 (1963),
{\it ibid.\/}, 1030. 
 H.LEUTWYLER, Nuovo.Cim. {bf 37}, 556 (1965)

\item {\dvqu}       Ph.DROZ-VINCENT, Found.of Phys. {\bf 25}, 67 (1995).

\item {\belfus}  L.BEL, X.FUSTERO,
 Mecanique relativiste predictive des systemes de N 
particules  Ann.Inst. Henri Poincar'e, {\bf A 25}, 411-436, (1976).

\item {\hj}  In relativistic mechanics
 HJ coordinates have been considered first in the two-body problem, with 
asymptotic conditions in time by 
L.BEL, J.MARTIN, Ann. Inst. Henri Poincar\'e, {\bf A 22}, 173-199 (1975). 
Later we advocated the use of HJ coordinates for the explicit construction of 
two-body (see  {\dvlxxvii}) and $n$-body interactions 
Ph.DROZ-VINCENT, C.R.Acad.Sc.Paris, {\bf 299 II}, 139 (1984).
Construction of relativistic interactions by a map to free motion was also
 considered without Hamiltonians by R.ARENS,
 contribution to ref. {\aadis}, p.1.

\item {\lapbar} X.BARCONS and R.LAPIEDRA, 
 Phys.Rev.  {\bf 28},  3030  (1983)
                    
\end